\documentclass[a4paper,11pt,oneside]{article}

\usepackage{amsmath}

\usepackage[english]{babel}
\usepackage[T1]{fontenc}
\usepackage[utf8]{inputenc}
\usepackage{newtxtext,newtxmath} 

\usepackage[pdftex,unicode]{hyperref}
\hypersetup{pdftitle=Modelling time-varying rankings with autoregressive and score-driven dynamics}
\hypersetup{pdfauthor=Vladimír Holý and Jan Zouhar}

\usepackage[margin=60pt]{geometry}
\DeclareMathOperator*{\plim}{plim}
\DeclareMathOperator{\PP}{P}
\DeclareMathOperator{\EE}{E}
\DeclareMathOperator{\lse}{lse}
\newcommand{\myexp}[1]{\ensuremath{\exp #1}}  
\newcommand{\rth}[1]{\ensuremath{{#1^\text{th}}}} 
\newcommand{\rthtilde}[1]{\ensuremath{{#1^\text{th}}}} 

\usepackage{enumitem}

\usepackage{graphicx}
\usepackage{xcolor}
\usepackage{float}
\usepackage{hhline}
\usepackage{multirow}
\usepackage{rotating}
\usepackage{booktabs}
\usepackage{dcolumn}
\usepackage{threeparttable}
\usepackage[format=hang, justification=raggedright, font=small, labelfont=bf, skip=0.5ex]{caption}
\usepackage[super]{nth}

\usepackage[authoryear]{natbib}

\hypersetup{colorlinks, allcolors={blue!50!black}}

\begin{document}

\begin{center}
{\Large \bfseries Modelling time-varying rankings with autoregressive and score-driven dynamics}
\end{center}

\begin{center}
{\bfseries Vladimír Holý} \\
Prague University of Economics and Business \\
Winston Churchill Square 1938/4, 130 67 Prague 3, Czechia \\
\href{mailto:vladimir.holy@vse.cz}{vladimir.holy@vse.cz}
\end{center}

\begin{center}
{\bfseries Jan Zouhar} \\
Prague University of Economics and Business \\
Winston Churchill Square 1938/4, 130 67 Prague 3, Czechia \\
\href{mailto:zouharj@vse.cz}{zouharj@vse.cz}
\end{center}

\begin{center}
{\itshape \today}
\end{center}

\begin{abstract}
\noindent We develop a new statistical model to analyse time-varying ranking data. The model can be used with a large number of ranked items, accommodates exogenous time-varying covariates and partial rankings, and is estimated via the maximum likelihood in a straightforward manner. Rankings are modelled using the Plackett-Luce distribution with time-varying worth parameters that follow a mean-reverting time series process. To capture the dependence of the worth parameters on past rankings, we utilise the conditional score in the fashion of the generalised autoregressive score (GAS) models. Simulation experiments show that the small-sample properties of the maximum-likelihood estimator improve rapidly with the length of the time series and suggest that statistical inference relying on conventional Hessian-based standard errors is usable even for medium-sized samples. In an empirical study, we apply the model to the results of the Ice Hockey World Championships. We also discuss applications to rankings based on underlying indices, repeated surveys, and non-parametric efficiency analysis.
\end{abstract}

\\

\\

\section{Introduction}
\label{sec:intro}

The rankings of universities, scientific journals, sports teams, election candidates, top-visited websites, or products preferred by customers are all examples of ranking data. Statistical models of ranking data have a long history, dating back at least to \cite{Thurstone1927}. Since then, the breadth of the statistical toolkit for ranking data has increased rapidly; see, e.g.,\ \cite{Marden1995} and \cite{Alvo2014} for an in-depth textbook overview. However, a recent survey of the ranking literature by \cite{Yu2019} draws attention to the lack of the time perspective in rankings and calls for research in this particular direction. This paper aims to heed this call and contribute to the thin strand of literature on time-varying ranking data. Unlike the existing models for time variation in rankings, our approach aims to provide a flexible tool for the modelling of time-varying ranking data that is similar to the autoregressive moving average (ARMA) model in the case of continuous variables.

Our model builds upon the (static) \emph{Plackett-Luce distribution} of \cite{Luce1959} and \cite{Plackett1975}, a convenient and simple probability distribution on rankings utilising a \emph{worth parameter} for each item to be ranked. It originates from Luce's \emph{choice axiom} and is also related to the Thurstone's \emph{theory of comparative judgment} (see \citealp{Luce1977} and \citealp{Yellott1977} for details). Although it is not without limitations, the Plackett-Luce distribution is widely used as a base for statistical models that are used to analyse ranking data. 

This also holds true for the scarce literature devoted to models with time-varying ranks. \cite{Baker2015} utilise the Plackett-Luce model and consider the individual worth parameters behind the rankings to be time-varying -- but deterministically so -- in an application to golf tournament results. \cite{Glickman2015} also base their model on the Plackett-Luce distribution but consider worth parameters following the Gaussian random walk in an application to women's alpine downhill skiing results. \cite{Asfaw2017} take a different path and include the lagged ranking as the current modal ranking in the Mallows model in an application to the academic performance of high school students. Finally, \cite{Henderson2018} employ the Plackett-Luce model with observations weighted in time in an application to Formula One results. The latter three papers adopt a Bayesian approach.

\emph{The generalised autoregressive score} (GAS) models of \cite{Creal2013}, which are also called \emph{dynamic conditional score} (DCS) models by \cite{Harvey2013}, have established themselves as a useful modern framework for time series modelling. The GAS models are observation-driven models allowing for any underlying probability distribution with any time-varying parameters. They capture the dynamics of time-varying parameters using the autoregressive term and the lagged score, i.e.,\ the gradient of the log-likelihood function. The GAS class includes many well-known econometric models, such as the \emph{generalised autoregressive conditional heteroskedasticity} (GARCH) model of \cite{Bollerslev1986}, which is based on the normal distribution with time-varying variance; the \emph{autoregressive conditional duration} (ACD) model of \cite{Engle1998}, which is based on the exponential distribution with a time-varying scale; and the count model of \cite{Davis2003}, which is based on the Poisson distribution with a time-varying mean. The GAS models can be straightforwardly estimated by the maximum likelihood method (see, e.g.,\ \citealp{Blasques2018} for details on the asymptotic theory). Generally, the GAS models perform very well when compared to alternatives (see, e.g.,\ \citealp{Koopman2016} for an extensive empirical and simulation study). Currently, the website \url{www.gasmodel.com} lists over 200 scientific papers devoted to the GAS models.

In the paper, we propose a dynamic model for rankings based on the Plackett-Luce distribution with time-varying worth parameters following the GAS score-driven dynamics. Our formulation allows for exogenous covariates and corresponds to the setting of a panel linear regression with fixed effects. We also consider the case of partial rankings. The proposed model is described in Section~\ref{sec:model}.

Using simulations, we investigate the finite-sample performance of the maximum likelihood estimator of our model. First, we demonstrate the convergence of the estimated coefficients for exogenous variables and of the GAS dynamics to their true values along the time dimension. Second, we show that confidence intervals based on the standard maximum likelihood asymptotics appear to be usable even if the dimensions of data are moderate (such as 20 items ranked in 20 time periods). The simulation study is conducted in Section~\ref{sec:sim}.

To demonstrate the proposed methodology, we analyse the results of the Ice Hockey World Championships from 1998 to 2019. We find that the proposed mean-reverting model fits the data better than the static and random walk models. The benefits of our approach include a compilation of the ultimate (long-term) ranking of teams, the straightforward estimation of the probabilities of specific rankings (e.g.,\ podium positions), and the prediction of future rankings. The empirical study is presented in Section~\ref{sec:hockey}.

Besides sports statistics, we discuss several other possible applications of the proposed model. Notably, we argue that our approach can be used to model rankings based on underlying indices (such as various country rankings) and captures the interaction between items, which the univariate models used directly for indices do not account for. Furthermore, we note that our model is suitable for repeated surveys that are designed as rankings. Finally, we show how our approach can be utilised for the rankings of decision-making units obtained by non-parametric efficiency analysis. These applications are discussed in Section~\ref{sec:disc}.

\section{Dynamic score-driven ranking model}
\label{sec:model}

\subsection{Plackett-Luce distribution}
\label{sec:modelDist}

Let us consider a set of $N$ items $\mathcal{Y} = \{1, \ldots, N \}$. Our main object of interest is a complete permutation of this set $y = \left( y(1), \ldots, y(N) \right)$, known as a \emph{ranking}, and its inverse $y^{-1} = \left( y^{-1}(1), \ldots, y^{-1}(N) \right)$, known as an \emph{ordering}. Element $y(i)$ represents the rank given to item $i$ while $y^{-1}(r)$ represents the item with rank $r$; to enhance readability, in subscripts we will simply write $\rth{r}$ instead of $y^{-1}(r)$ to denote the item ranked \rth{r}. 

We assume that a random permutation $Y$ follows the \emph{Plackett-Luce distribution} of \cite{Luce1959} and \cite{Plackett1975}. According to this distribution, a ranking is constructed by successively selecting the best item, the second best item, the third best item, and so on. The probability of selecting a specific item in any stage is equal to the ratio of its worth parameter and the sum of the worth parameters of all items that have not yet been selected. Therefore, the probability of a complete ranking $y$ is
\begin{equation}
\label{eq:luceCompProb}
	\PP \left[ Y = y \middle| f \right] = \prod_{r=1}^{N} \frac{\myexp{f_\rth{r}}}{ \sum_{s=r}^N \myexp{f_\rth{s}} },
\end{equation}
where $f = (f_1,\ldots,f_N)'$ are the items' worth parameters. We use a parametrization allowing for arbitrary values of $f_i$, which facilitates subsequent modelling. Note that the probability mass function \eqref{eq:luceCompProb} is invariant to the addition of a constant to all parameters $f_i$. Therefore, we employ the standardisation
\begin{equation}
	\label{eq:luceStan}
	\sum_{i=1}^N f_i = 0.
\end{equation}

The log-likelihood function is 
\begin{equation}
\label{eq:luceCompLik}
\ell \left( f \middle| y \right) = \sum_{i=1}^{N} f_i - \sum_{r=1}^{N} \ln \left( \sum_{s=r}^N \myexp{f_\rth{s}} \right).
\end{equation}
For a random sample of rankings, a necessary and sufficient condition for the log-likelihood to have a unique maximum is that in every possible partition of $\mathcal{Y}$ into two non-empty subsets, some item in the second set ranks higher than some item in the first set at least once \citep{Hunter2004}. This condition, for example, rules out that there is an item always ranked first (in maximum likelihood estimation, this would result in an infinite worth parameter). To overcome the limitations of this condition in practical applications, \citet{Luo2019} propose a penalised maximum likelihood estimator that adds a small perturbation to the log-likelihood.

For further details regarding the Plackett-Luce distribution, see \cite{Luce1977}, \cite{Yellott1977}, \cite{Stern1990}, and \cite{Critchlow1991}.

\subsection{Conditional score}
\label{sec:modelScore}

The key ingredient in our dynamic model is the score, i.e.,\ the gradient of the log-likelihood function, which is defined as
\begin{equation}
\label{eq:defScore}
\nabla \left( f \middle| y \right) = \frac{\partial \ell \left( f \middle| y \right)}{\partial f}.
\end{equation}
The score represents the direction for improving the fit of the distribution with a given $f$ to a specific observation $y$ and indicates the sensitivity of the log-likelihood to the parameter $f$. For a complete ranking $y$ following the Plackett-Luce distribution, the score is given by
\begin{equation}
\label{eq:luceCompScore}
\nabla_i \left( f \middle| y \right) = 1 - \sum_{r=1}^{y(i)} \frac{\myexp{f_i}}{\sum_{s=r}^N \myexp{f_\rth{s}} }, \qquad i = 1, \ldots, N.
\end{equation}
An example with three items, in which the score is easily obtained, is given in Appendix~\ref{sec:threeItems}. Appendix~\ref{sec:softmax} rewrites the score formula using the softmax function and shows additional steps in its derivation. In general, the score function has zero expected value and its variance is equal to the Fisher information:
\begin{equation}
\label{eq:defFisher}
\mathcal{I}(f) = \EE \left[ \nabla \left( f \middle| y \right) \nabla \left( f \middle| y \right)' \middle| f \right].
\end{equation}
Although the Fisher information is available in a closed form for the Plackett-Luce distribution, it is computationally very intensive for larger $N$ as it includes a sum over all possible permutations of $\mathcal{Y}$.

The score has an appealing interpretation. In essence, it reflects the discrepancy between the items' worth parameters and the eventual ranking. This information can be exploited in a time-series context where the worth parameters are updated with each new observation. For example, consider a tournament with teams $A$, $B$, and $C$ with worth parameters $f = (2, 0, -2)'$; here, $f_i$ can be interpreted as a measure of team $i$'s strength. If the tournament goes as expected and the order of the players is $(A, B, C)$ -- which happens with a probability of 76.3\% -- the score is close to zero for all players: $\nabla \left[ f \middle| (A, B, C) \right] = (0.13, 0.0019, -0.14)'$. However, if the order is reversed (an outcome occurring with a probability of a mere 0.2\%), the score is  $\nabla \left[ f \middle| (C, B, A) \right] = (-1.75, 0.76, 0.98)'$. For team $A$, which failed despite the high expectations, the score is negative; for those who beat $A$, the score is positive, with the largest score obtained for the unlikely winner $C$. The score can therefore serve as a basis for the correction of worth parameters after an observation is realised.

\begin{figure}
    \begin{center}
        \includegraphics[width=0.8\textwidth]{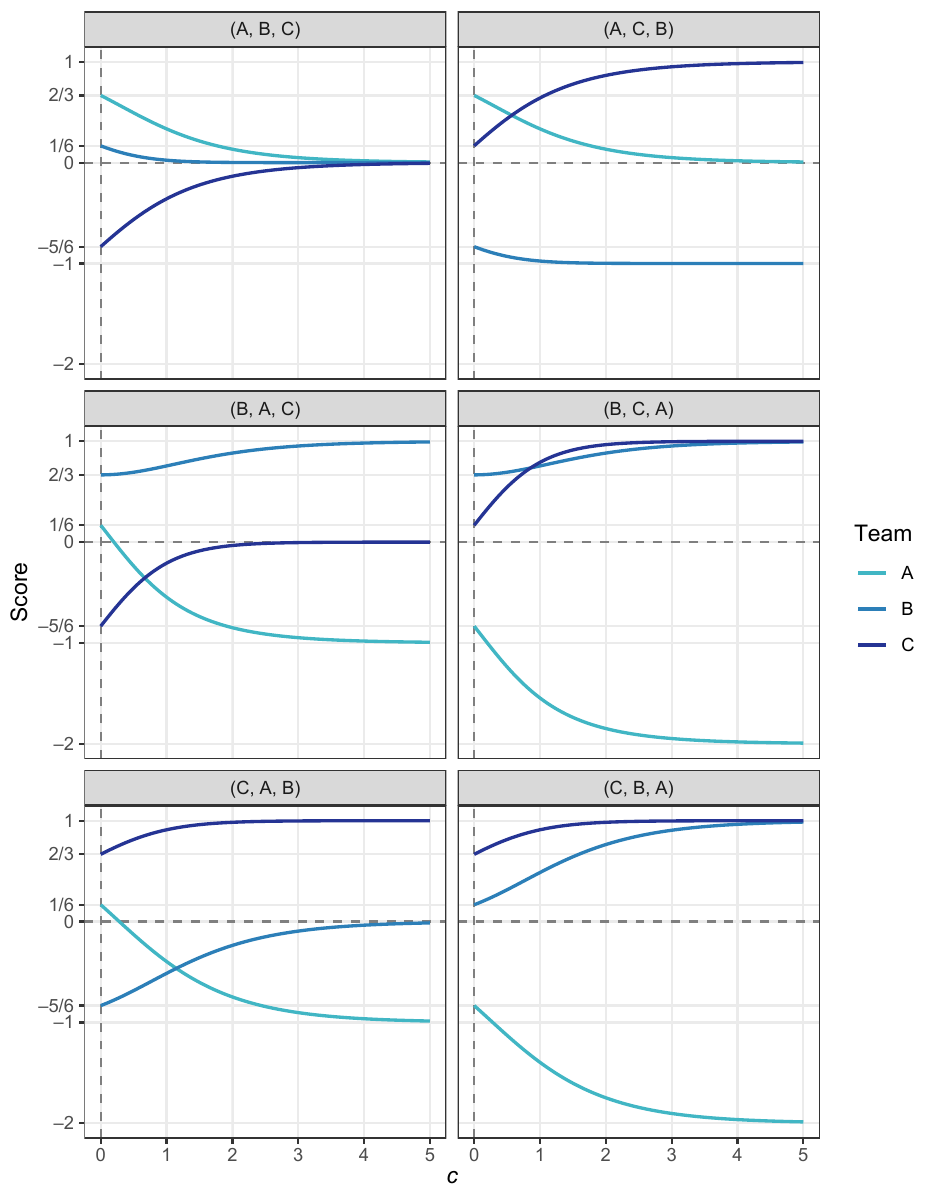}
        \caption{Scores in the Plackett-Luce distribution for three items ($A$, $B$, and $C$), worth parameters $f = (c, 0, -c)'$ for $c\in[0,5]$ (horizontal axis), and all possible orderings (panel titles).}
        \label{fig:figScore}
    \end{center}
\end{figure}

Figure~\ref{fig:figScore} extends the previous example: it shows the score for $f = (c, 0, -c)'$ at different levels of $c>0$ under all six orderings. For large values of $c$, the score appears to converge to integer values. This is no coincidence: the score is bounded by integer values. As we show in Appendix~\ref{sec:softmax}, in the general case of $N$ items, the score always lies in $(1 - r, 1)$ for the item with rank $r=1,\ldots,N-1$ and in $(1-N, 0)$ for the item with rank $N$.

\subsection{Score-driven dynamics}
\label{sec:modelDyn}

Let us observe the rankings $y_t$ in times $t=1, \ldots, T$. Furthermore, let us assume that individual worth parameters evolve over time and denote them $f_t = (f_{1,t}, \ldots, f_{N,t})'$ for $t = 1, \ldots, T$. Specifically, let the time-varying parameter $f_{i,t}$ follow the generalised autoregressive score (GAS) dynamics of \cite{Creal2013} and \cite{Harvey2013} with a score order $P$ and an autoregressive order $Q$. Let it also linearly depend on exogenous covariates $x_1, \ldots, x_M$. The parameter $f_{i,t}$ is then given by the recursion
\begin{equation}
\label{eq:gasModel}
f_{i, t} = \omega_i + \sum_{j=1}^M \beta_{j} x_{i,t,j} + \sum_{k=1}^P \alpha_{k} \nabla_i \left( f_{t - k} \middle| y_{t - k} \right) + \sum_{l=1}^Q \varphi_{l} f_{i,t - l}, \qquad i = 1, \ldots, N, \quad t = 1, \ldots, T,
\end{equation}
where $\omega_i$ is item $i$'s individual fixed effect, $\beta_j$ is the regression parameter on $x_j$, $\alpha_k$ is the score parameter for lag $k$, $\varphi_l$ is the autoregressive parameter for lag $l$, and $x_{i,t,j}$ is the value of $x_j$ for item $i$ at time $t$. Note that this formulation corresponds to the setting of a panel linear regression with fixed effects. In most of the GAS literature (and the GARCH and ACD literature, as a matter of fact), only the first lags are utilised, i.e.,\ $P=Q=1$.

In the GAS framework, the score function can be scaled by the inverse of the Fisher information or the inverse of the square root of the Fisher information, although the unit scaling is often utilised as well (see \citealp{Creal2013}). The right choice of scaling can make estimators robust by mitigating the effect of outlying realisations of $y_t$ on time-varying parameters. A well-known case is the Beta-t-GARCH model of \cite{Harvey2008}, the GAS counterpart to Bollerslev's \citeyearpar{Bollerslev1987} GARCH-t model: by applying the inverse-information scaling in Beta-t-GARCH, one obtains a model with a milder response of the variance to a large $|y_t|$ than that in  GARCH-t \citep[Ch.~4]{Harvey2013}. In the case of the Plackett-Luce distribution, analogous robustness properties are already in place with unit scaling (i.e., no scaling) thanks to the boundedness of the Plackett-Luce score. In fact, it turns out that inverse-information scaling will typically make the effect of outlying observations on the worth parameters more pronounced. Moreover, obtaining the Fisher information is computationally very intensive even for moderate $N$, as it involves a sum over $N!$ permutations. For these reasons, we only consider unit scaling.

Standardisation \eqref{eq:luceStan} cannot be enforced at each time $t$ without deforming the dynamics given by the recursion \eqref{eq:gasModel}. Instead, we use the standardisation
\begin{equation}
\label{eq:gasStan}
\sum_{i=1}^N \omega_i = 0.
\end{equation}
In the case of mean-reverting dynamics with no exogenous covariates, this corresponds to a zero sum of the unconditional values:
\begin{equation}
\label{eq:gasUnc}
\bar{f}_i = \frac{\omega_i}{1 - \sum_{l=1}^Q \varphi_{l}}, \qquad i=1, \ldots, N.
\end{equation}

\subsection{Maximum likelihood estimation and inference}
\label{sec:modelEst}

For the estimation of the proposed dynamic model, we utilise the maximum likelihood estimator. Let $\theta = (\omega_1, \ldots, \omega_{N-1}, \beta_1, \ldots, \beta_M, \alpha_1, \ldots, \alpha_P, \varphi_1, \ldots, \varphi_Q)'$ be the vector of the $N+M+P+Q-1$ parameters to be estimated, with $\omega_N$ being obtained from \eqref{eq:gasStan} as $\omega_N = - \sum_{i=1}^{N-1}\omega_i$. The estimate $\hat{\theta}$ is obtained from the conditional log-likelihood as
\begin{equation}
\label{eq:mle}
\hat{\theta} \in \arg\max_\theta \sum_{t=1}^T \ell \left( f_t \middle| y_t \right).
\end{equation}
The recursive nature of $f_t$ requires the initialisation of the first few elements of the conditional score and worth parameter time series. A reasonable approach is to set the initial conditional scores $\nabla(f_0 | y_0)$, \ldots, $\nabla(f_{-P+1} | y_{-P+1})$ to zero, i.e.,\ their expected value, and the initial parameters $f_0$, \ldots, $f_{-Q+1}$ to the unconditional value $\bar{f}$ given by \eqref{eq:gasUnc}. Alternatively, if additional information about the initial worth parameters is available, it can be used instead. For instance, in a related GAS-type model for the binary outcomes of tennis matches, \cite{Gorgi2019} use current ranking points to initialise the worth parameters. On the other hand, they also note that other initialisation methods yielded very similar parameter estimates. The initial worth parameters can also be considered as additional parameters to be estimated. This would, however, significantly increase the number of variables in the maximisation problem.

From a computational perspective, it is possible to utilise any general-purpose algorithm to solve nonlinear optimisation problems. In our simulation study and empirical application, we employ the \emph{Broyden--Fletcher--Goldfarb--Shanno} (BFGS) algorithm. The optimisation performance can be improved, however, by exploiting the special structure of the problem. Concerning the GAS models, \cite{Creal2013} recall the work of \cite{Fiorentini1996} and suggest an algorithm that computes the gradient of the likelihood recursively and simultaneously with the time-varying parameters. Concerning the Plackett-Luce distribution, \cite{Hunter2004} presents an iterative \emph{minorization–maximization} (MM) algorithm, which is further developed by \cite{Caron2012}. These ideas might prove to be a useful starting point for a specialised likelihood-maximisation algorithm tailored to our model; however, the development of such an algorithm is beyond the scope of this paper.

Our implementation of statistical inference tasks is based on standard maximum likelihood asympotics. Recall that under suitable regularity conditions, the maximum likelihood estimator $\hat{\theta}$ is consistent and asymptotically normal, i.e.,\ it satisfies
\begin{equation}
\sqrt{T} \big( \hat{\theta} - \theta_0 \big) \stackrel{\mathrm{d}}{\to} \mathrm{N} \big( 0, -H^{-1} \big), 
\end{equation}
where $H$ denotes the asymptotic Hessian of the log-likelihood, defined as
\begin{equation}
H = \plim_{T \to \infty} \frac{1}{T} \sum_{t=1}^T 
      \frac{\partial^2 \ln \PP \left[ Y_t = y_t \middle| f_t \right]}
            {\partial \theta_0 \partial \theta_0'}.
\end{equation}
In finite samples, standard errors are often computed using the empirical Hessian of the log-likelihood evaluated at $\hat\theta$, and the normal c.d.f. is used for statistical inference.

The truth is that establishing the asymptotic theory for GAS-type models is difficult in general. At a minimum, it is necessary that the filter $f_t$ is invertible (see, e.g.,\ \citealp{Blasques2018} for more details). The invertibility property ensures, among other things, that the initialisation does not matter in the long run. The theoretical derivation of the conditions restricting the parameter space in order to obtain consistency and asymptotic normality is, however, beyond the scope of this paper. Indeed, it is very challenging in general to obtain any asymptotic results for the case of multivariate variables with multiple time-varying parameters. In the following, we base our inference on the asymptotics outlined above and rely on simulations to verify their validity.

\subsection{Extension to partial rankings}
\label{sec:modelPart}

The distribution can be extended to the case in which the ranking of only the top $\tilde{N} < N$ items is observed. The set of ranked items is then $\tilde{\mathcal{Y}} = \{ y^{-1}(1), \ldots, y^{-1}(\tilde{N}) \}$. We denote the partial ranking of items $i \in \tilde{\mathcal{Y}}$ as $\tilde{y}$ and the partial ordering as $\tilde{y}^{-1}$. The probability mass function of the Plackett-Luce distribution for the partial ranking $\tilde{y}$ is then
\begin{equation}
\label{eq:lucePartProb}
\PP \left[ \tilde{Y} = \tilde{y} \middle| f \right] = \prod_{r=1}^{\tilde{N}} \frac{\myexp{f_\rth{r}}}{ \sum_{s=r}^{\tilde{N}} \myexp{f_\rth{s}} + \sum_{j \not\in \tilde{\mathcal{Y}}} \myexp{f_j} }.
\end{equation}
The score function for the partial ranking $\tilde{y}$ is
\begin{equation}
\label{eq:lucePartScore}
\nabla_i \left( f | \tilde{y} \right)  =
\begin{cases}
1 - \sum_{r=1}^{\tilde{y}(r)} \frac{\myexp{f_i}}{\sum_{s=r}^{\tilde{N}} \myexp{f_\rthtilde{s}} + 
	\sum_{j \not\in \tilde{\mathcal{Y}}} \myexp{f_j} } 
	& \text{for } i \in \tilde{\mathcal{Y}}, \\[3ex]
- \sum_{r=1}^{\tilde{N}} \frac{\myexp{f_i}}{\sum_{s=r}^{\tilde{N}} \myexp{f_\rthtilde{s}} + 
    \sum_{j \not\in \tilde{\mathcal{Y}}} \myexp{f_j} } 
	& \text{for } i \not\in \tilde{\mathcal{Y}}. \\
\end{cases}
\end{equation}
The probability mass function and the score function for partial rankings can be straightforwardly plugged into the dynamics from previous sections.

\section{Finite-sample performance}
\label{sec:sim}

\subsection{Simulation design}
\label{sec:simDesign}

We conduct a simulation study in order to investigate the behaviour of the maximum likelihood estimator over two dimensions -- the number of items $N$ and the time horizon $T$. In particular, $N$ varies between 10, 20, and 30, and $T$ ranges from 10 to 100. For each combination of $N$ and $T$, we conduct 100,000 replications.

The simulations employ the following toy model with the parameters selected to resemble those estimated in the empirical study in Section \ref{sec:hockey}. As we consider different values of $N$, the number of $\omega_i$ parameters differs. To somewhat standardise the item-specific fixed effects, we set $\omega_i = 4 (i - 1) / (N - 1) - 2$, $i=1,\ldots,N$, i.e.,\ the parameters $\omega_i$ range from $-2$ to $2$ for any $N$. We include a single exogenous covariate independently generated from the standard normal distribution. The regression parameter is then set to $\beta_1 = 1$. Finally, the order of the GAS model is chosen as $P = Q = 1$, with the dynamics parameters set to $\alpha_1 = 0.4$ and $\varphi_1 = 0.5$. Such parameter values result in the unconditional values $\bar{f}_i$ given by \eqref{eq:gasUnc}, which range from $-4$ to $4$. In the following, we drop unnecessary subscripts for brevity, and simply refer to $\beta_1$ as $\beta$, $\alpha_1$ as $\alpha$, and $\varphi_1$ as $\varphi$.

\subsection{Simulation results}
\label{sec:simResults}

\begin{figure}
	\begin{center}
		\includegraphics[width=0.9\textwidth]{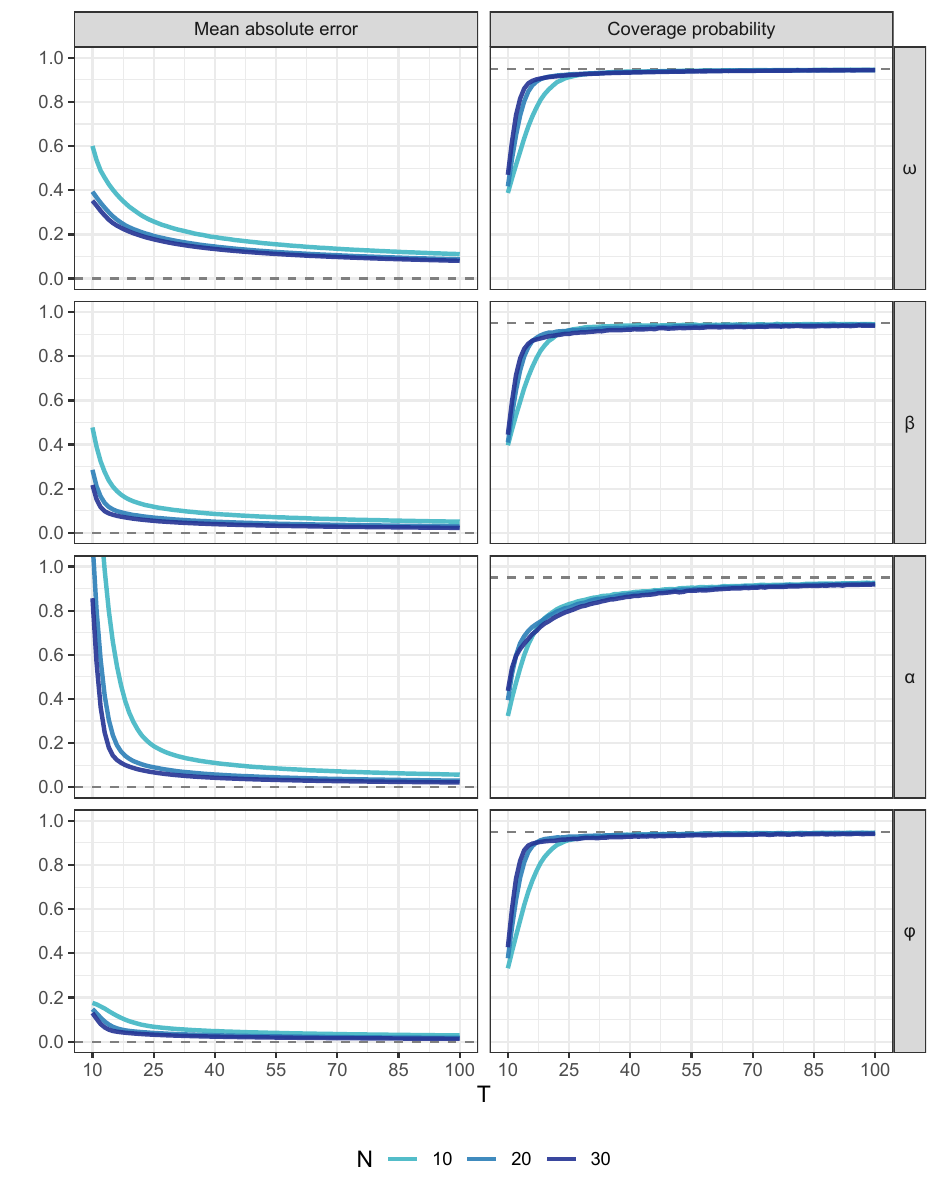}
		\caption{Mean absolute errors of the estimated coefficients and coverage probabilities of the 95\% confidence intervals. For the item-specific fixed effects, $\hat\omega_i$, the results are averaged across all items. Dashed horizontal lines are drawn at the 0 and 0.95 vertical coordinates to show the limit values of mean absolute errors and the coverage probabilities under standard maximum-likelihood asymptotics.}
		\label{fig:figSimError}
	\end{center}
\end{figure}

The results of the simulation study are reported in Figure \ref{fig:figSimError}. First, we investigate the accuracy of the estimators $\hat{\omega}_i$, $\hat{\beta}$, $\hat{\alpha}$, and $\hat{\varphi}$; to enhance readability, the results for $\hat{\omega}_i$ are averaged across all items ($i$). In the first column of Figure \ref{fig:figSimError}, we report the mean absolute errors (MAE) between the estimated coefficients and their true values. All estimates converge to their true values along the time dimension. The score parameter $\alpha$ proves to be hard to estimate for small $T$ as it has much higher MAE than the autoregressive parameter $\varphi$ with a comparable nominal value. Nevertheless, even in a medium sample with $N=20$ and $T=20$, the errors are not that substantial, with values of 0.22 for $\hat{\omega}_i$, 0.08 for $\hat{\beta}$, 0.12 for $\hat{\alpha}$, and 0.05 for $\hat{\varphi}$. In a large sample with $N=30$ and $T=100$, the errors further decrease to 0.08 for $\hat{\omega}_i$, 0.02 for $\hat{\beta}$, 0.02 for $\hat{\alpha}$, and 0.01 for $\hat{\varphi}$.

Second, we assess the usability of the maximum likelihood asymptotics for finite-sample inference. Specifically, in the second column of Figure \ref{fig:figSimError}, we report the fraction of the samples in which the 95 percent confidence intervals contained the true parameter values, i.e.,\ we present the estimated coverage probabilities. For all parameters, the coverage probability converges to the target value of 0.95 from below along the time dimension. As in the case of MAE, the convergence of the coverage probability is the slowest for the score parameter $\alpha$. In a medium sample with $N=20$ and $T=20$, the coverage probabilities are 0.91 for $\hat{\omega}_i$, 0.91 for $\hat{\beta}$, 0.78 for $\hat{\alpha}$, and 0.92 for $\hat{\varphi}$, while in a large sample with $N=30$ and $T=100$, they amount to 0.94 for $\hat{\omega}_i$, 0.94 for $\hat{\beta}$, 0.92 for $\hat{\alpha}$, and 0.94 for $\hat{\varphi}$.

\section{Application to ice hockey rankings}
\label{sec:hockey}

\subsection{Data set}
\label{sec:hockeyData}

We demonstrate the use of our model using data on the results of the Ice Hockey World Championships between the years 1998 and 2019. In 1998, the sanctioning body of the championships, the International Ice Hockey Federation (IIHF), increased the number of teams in the tournament from 12 to 16, and has kept the number of teams at that level since then; hence, 1998 was chosen as the starting year. For each year, the IIHF provides a complete ranking of all 16 participants. Over the years, 24 different teams made it through the qualification process, and they comprise the set of ranked items in our model.

For each year, we obtained information about the host country of the championships. In order to account for the home-ice advantage, we included a \emph{home ice} covariate, which is a time-varying indicator variable (it is equal to 1 for home teams in the respective years, and it is equal to 0 otherwise).

\subsection{Model specification}
\label{sec:hockeyModel}

The general structure of the team strength dynamics given by \eqref{eq:gasModel} includes an array of different model specifications that can be obtained by (i) choosing the order of the GAS model ($P, Q$) and (ii) imposing specific restrictions on the parameter space. As for the former, with the limited size of our data set, it seems impractical to consider anything beyond the canonical $P = Q = 1$ model.

Setting $P = Q = 0$, on the other hand, yields a static strength model, which is equivalent to the standard \emph{ranked-order logit} (ROL) -- a common go-to model for sports rankings. Recent applications to sport rankings include, e.g.,\ \cite{Caron2012} and \cite{Henderson2018}. The latter use time-weighted observations to improve forecasts, but their model is intrinsically static. Both referenced studies use a Bayesian approach to the estimation of the ROL model. We estimate the \textbf{static model} to provide a benchmark for the models with score-driven dynamics.

In the model with $P = Q = 1$, we generally expect the autoregressive parameter to lie in the $(0, 1)$ interval, implying a certain degree of persistence in team strengths with a mean-reverting tendency. In our data set, this is indeed the result we obtain if we leave the parameters unrestricted in the likelihood-maximisation procedure. We refer to this variant as the \textbf{mean-reverting model}. The need for this type of a sports ranking model has recently been recognised by \cite{Baker2015}. In their analysis of golf tournament rankings, they note that while their model's deterministic dynamics do sufficiently capture the time variation in an individual player's performance, a mean-reverting random process would be more appropriate for teams. The performance of individual players tends to follow long-term trends, potentially with breakpoints (due to injuries, the long-term evolution of self-confidence, ageing, etc.). Teams, on the other hand, do not have a fixed membership structure; players come and go on a relatively flexible basis, depending on their current performance. Massive exogenous shocks with a persistent effect are less common.

A GAS setting similar to ours has recently been used in the context of sports statistics by \cite{Koopman2019}. Rather than dealing with ranking data, Koopman and Lit focus on individual football matches, modelling either the qualitative (win-draw-loss) outcomes or match scores. Their estimates indicate high persistence levels in strength dynamics, with a typical value of (the equivalent to our) $\hat\varphi$ of around 0.998 for back-to-back matches in most of the estimated models. The authors note that this corresponds to a yearly persistence of about 0.90.

If strong persistence is expected, it might be reasonable to restrict the autoregressive parameter to unity, making the team strengths follow a random walk pattern. A ranking model of this type was presented by \cite{Glickman2015} in the context of women's alpine downhill skiing competitions. Their model, however, does not make use of the score-driven component and is estimated in a Bayesian framework. In a GAS setting, a model with a random walk behaviour of the team strength (henceforth, \textbf{random walk model}) should be approached with caution. The $f_t$ filter in this case is not invertible, making the consistency of the maximum likelihood estimator dubious. That said, in our simulation experiments, the mean absolute error of the coefficient estimates was roughly comparable to the mean-reverting model of equal sample size, provided that the model specifications agree with the underlying data-generating processes. We also note that a GAS model with random walk strength dynamics has recently been presented by \cite{Gorgi2019} to predict the outcomes of individual tennis matches. Gorgi et al.\ view mean-reverting processes as the dynamics of choice for team sports and non-stationary dynamic processes as more appropriate for individual sports.

Using the general framework developed in Section~\ref{sec:model}, we can easily estimate all three model variants (the static model, mean-reverting model, and random walk model) and compare their fits using information-theoretic criteria. As only a subset of all teams participated in each championship, we employ the form of the likelihood function for partial rankings developed in Section~\ref{sec:modelPart}.

The computation is performed using R package \verb"gasmodel" for estimation, forecasting, and simulation of GAS models based on various distributions including the Plackett-Luce distribution. The package includes the analyzed Ice Hockey World Championships data and a vignette describing our modelling approach. It is available at \url{https://github.com/vladimirholy/gasmodel}.

\subsection{Empirical results}
\label{sec:hockeyRes}

In the observed period, 1998--2019, only 9 teams participated in all 22 Ice Hockey World Championships. These included all the teams from the so-called Big Six (Canada, Czechia, Finland, Russia, Sweden, and the United States) along with Latvia, Slovakia, and Switzerland. Three teams -- Great Britain, Poland, and South Korea -- only appeared once. The dominance of the Big Six is evident when looking at the podium positions: out of the 66 medals, only six were handed out to teams outside the Big Six (four were awarded to Slovakia and two to Switzerland). Hosting was also unevenly distributed among the countries: only 14 of the teams experienced the home-ice advantage, with Czechia, Slovakia, and Switzerland hosting the championships twice and Germany, Finland, Russia, and Sweden hosting them three times each.

Table \ref{tab:HockeyCoeffs} presents the results for all three estimated models. In terms of the Akaike information criterion (AIC), the mean-reverting model outperformed the remaining two by a wide margin, with $\Delta$AIC exceeding 25 in both cases. This (i) implies that the introduction of strength dynamics can improve the model fit dramatically and (ii) provides empirical support for the conjecture of \cite{Gorgi2019} about the suitability of mean-reverting dynamics for team sports. The 95\% confidence interval for the autoregressive parameter in the mean-reverting model, $[0.21, 0.80]$, indicates the presence of moderate persistence and leads us to reject the null hypotheses of both a random-walk behaviour and no serial dependence.

\begin{table}
\caption{Selected estimates for the Ice Hockey World Championships data (1998--2019).}
\label{tab:HockeyCoeffs}

\begin{threeparttable}
\begin{tabular}{l D{.}{.}{4.6} D{.}{.}{4.4} D{.}{.}{4.6}}
\toprule
 & \multicolumn{1}{c}{~~Mean-reverting model~~} & \multicolumn{1}{c}{~~Static model~~} & \multicolumn{1}{c}{~~Random walk model~~} \\
\midrule
Home ice ($\hat\beta$)                                 & 0.227       & 0.171    & 0.099       \\
                                                       & (0.258)     & (0.262)  & (0.188)     \\
\addlinespace Score parameter ($\hat\alpha$)           & 0.392^{***} &          & 0.343^{***} \\
                                                       & (0.083)     &          & (0.058)     \\
\addlinespace Autoregressive parameter ($\hat\varphi$) & 0.506^{***} &          &             \\
                                                       & (0.149)     &          &             \\
\midrule
log-likelihood                                         & -611.195    & -625.800 & -625.425    \\
AIC                                                    & 1274.391    & 1299.600 & 1300.851    \\
\bottomrule
\end{tabular}
\begin{tablenotes}[flushleft]
\scriptsize{\item[\hspace{-5mm}] \textit{Notes}: (i)~Estimates of $\omega_i$ are omitted from the table to enhance readability. (ii)~Standard errors in parentheses. (iii)~$^{***}p<0.001$; $^{**}p<0.01$; $^{*}p<0.05$.}
\end{tablenotes}
\end{threeparttable}

\end{table}

Despite the differences in AIC values, for parameters that are shared across the models, the estimates are qualitatively similar. In both the mean-reverting and random walk model, the values of the score coefficient, $\hat\alpha$, are positive and significant. This implies that the score component of our model does help in explaining the ranking dynamics. A positive sign of $\hat\alpha$ is in line with the interpretation of the conditional score outlined in Section~\ref{sec:modelScore}: a surprising success will positively affect the team's strength estimate for the next season and vice versa.

In accordance with expectations, point estimates in all three models suggest the existence of a home-ice advantage ($\hat\beta > 0$), but the \textit{home ice} is not statistically significant in either model.  To assess the effect size implied by $\hat\beta$, we need to know the team strengths. Table~\ref{tab:HockeyStrengths} shows the estimates of the unconditional team strength from the mean-reverting model, which are obtained based on \eqref{eq:gasUnc}, and the $\hat\omega_i$ for the static model. The differences in successive strength values suggest that an increase of $0.23$ (the home-ice advantage estimate in the mean-reverting model) moves a team 0--2 places ahead in the ranking.

\begin{table}
	\caption{Unconditional team strength estimates and ultimate ranking in the mean-reverting and static models. Teams are sorted by the ultimate ranking obtained from the mean-reverting model.}
	\label{tab:HockeyStrengths}
	
\begin{tabular}[t]{>{\raggedright\arraybackslash}p{2.5cm}>{\raggedleft\arraybackslash}p{1.8cm}>{\raggedleft\arraybackslash}p{1.8cm}>{\raggedleft\arraybackslash}p{1.8cm}>{\raggedleft\arraybackslash}p{1.8cm}}
\toprule
\multicolumn{1}{c}{ } & \multicolumn{2}{c}{Mean-reverting model} & \multicolumn{2}{c}{Static model} \\
\cmidrule(l{3pt}r{3pt}){2-3} \cmidrule(l{3pt}r{3pt}){4-5}
Country & Strength & Rank & Strength & Rank\\
\midrule
Canada & $3.72$ & 1 & $3.72$ & 2\\
Finland & $3.70$ & 2 & $3.66$ & 3\\
Sweden & $3.65$ & 3 & $3.84$ & 1\\
Czechia & $3.47$ & 4 & $3.41$ & 4\\
Russia & $3.25$ & 5 & $3.17$ & 5\\
United States & $1.83$ & 6 & $2.18$ & 6\\
\addlinespace
Switzerland & $1.67$ & 7 & $1.76$ & 7\\
Slovakia & $1.65$ & 8 & $1.55$ & 8\\
Latvia & $0.86$ & 9 & $0.82$ & 9\\
Germany & $0.28$ & 10 & $0.31$ & 10\\
Belarus & $0.25$ & 11 & $0.11$ & 11\\
Norway & $0.03$ & 12 & $-0.07$ & 12\\
\addlinespace
Denmark & $-0.07$ & 13 & $-0.17$ & 13\\
France & $-0.41$ & 14 & $-0.51$ & 14\\
Austria & $-0.83$ & 15 & $-0.89$ & 15\\
Italy & $-1.02$ & 16 & $-1.10$ & 16\\
Ukraine & $-1.34$ & 17 & $-1.52$ & 17\\
Slovenia & $-1.75$ & 18 & $-1.64$ & 18\\
\addlinespace
Kazakhstan & $-1.83$ & 19 & $-1.78$ & 19\\
Japan & $-2.00$ & 20 & $-1.94$ & 20\\
Hungary & $-3.28$ & 21 & $-3.20$ & 21\\
Great Britain & $-3.92$ & 22 & $-3.89$ & 22\\
Poland & $-3.95$ & 23 & $-3.90$ & 23\\
South Korea & $-3.96$ & 24 & $-3.91$ & 24\\
\bottomrule
\end{tabular}
\end{table}

For the mean-reverting and static models, the estimates of $\omega_i$ can be used to provide the `ultimate' (or long-run) ranking. Both models confirm the dominance of the Big Six. Indeed, the rankings in both models agree in all but the first three places; the long-term strength estimates for these three teams are very close to one another, though, making the eventual ranking less clear cut.

\begin{figure}
	\begin{center}
		\includegraphics[width=1.0\textwidth]{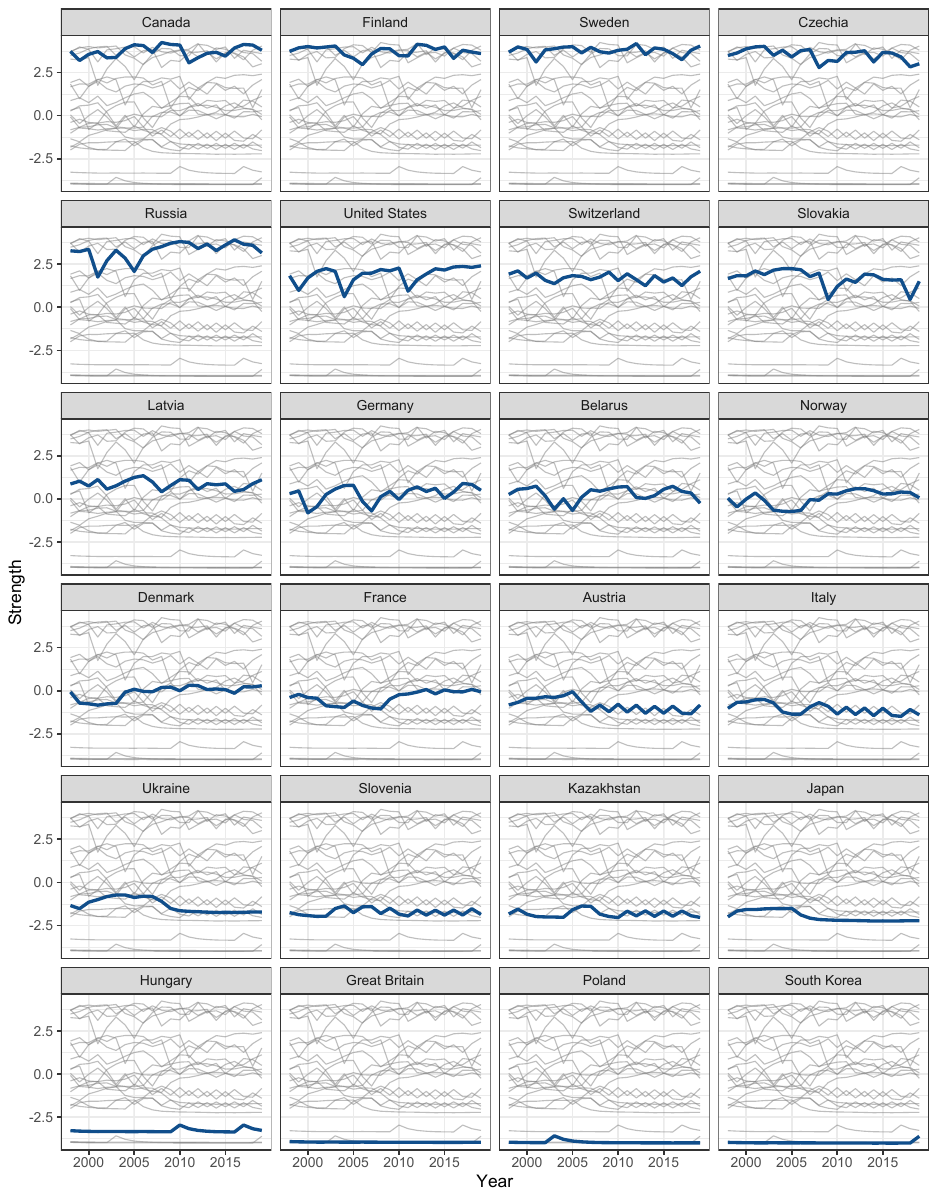}
		\caption{Mean-reverting model -- estimated team strength for all teams over the entire observed period. Teams are ordered by the estimated unconditional ranking.}
		\label{fig:figHockeyStrengthStationary}
	\end{center}
\end{figure}

\begin{figure}
	\begin{center}
		\includegraphics[width=1.0\textwidth]{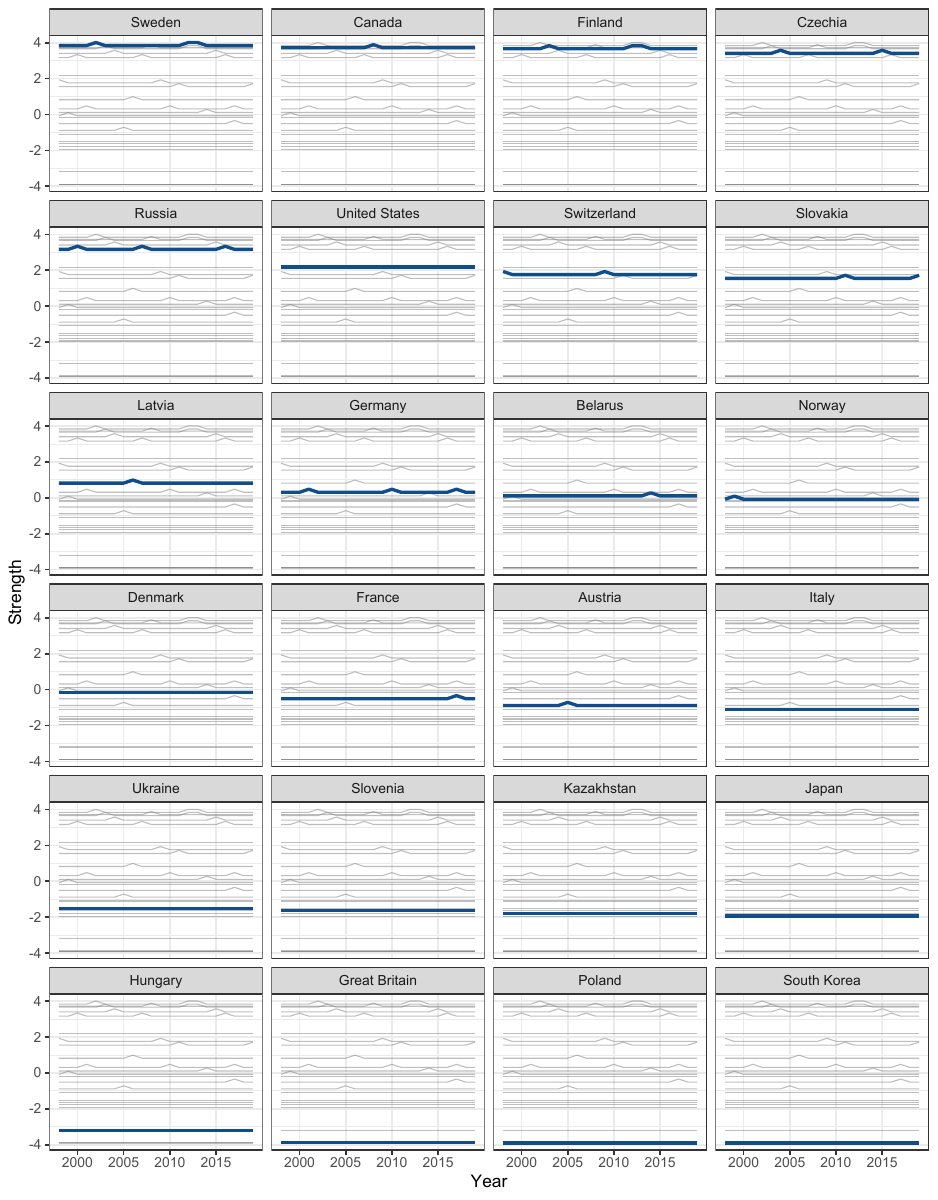}
		\caption{Static model -- estimated team strength for all teams over the entire observed period. Teams are ordered by the estimated unconditional ranking. In the static model, team strengths only vary with the home-ice advantage, producing little bumps in the plots.}
		\label{fig:figHockeyStrengthStatic}
	\end{center}
\end{figure}

\begin{figure}
	\begin{center}
		\includegraphics[width=1.0\textwidth]{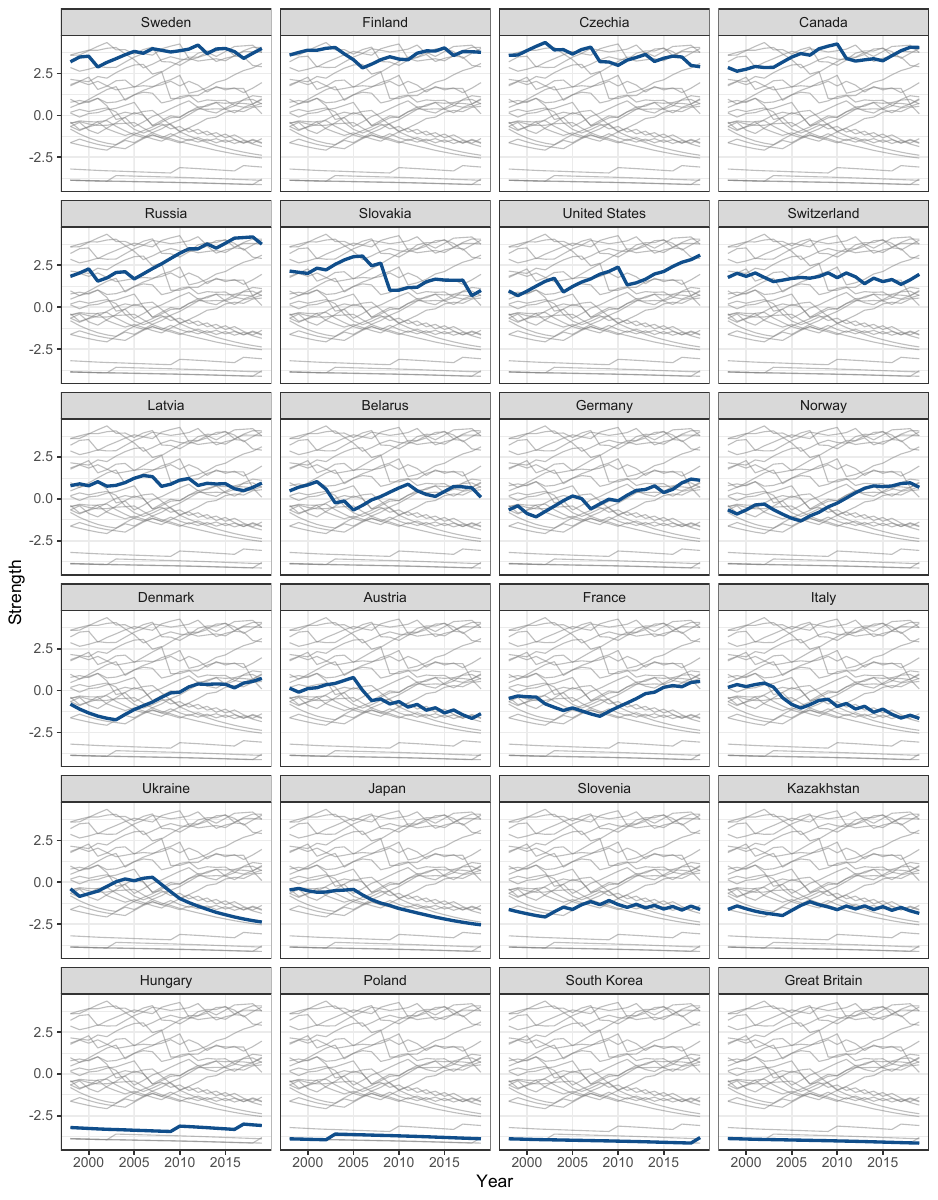}
		\caption{Random walk model -- estimated team strength for all teams over the entire observed period. Teams are ordered by the mean strength estimate.}
		\label{fig:figHockeyStrengthRandomWalk}
	\end{center}
\end{figure}

Figure~\ref{fig:figHockeyStrengthStationary} presents the estimated values of the worth parameters $f_{i,t}$ (referred to here as the strength) in the mean-reverting model. Even though the serial dependence, given by the autoregressive parameter $\alpha$, is mild, it is clearly discernible in the plots. For teams that only appeared in a handful of championships (i.e.,\ the weaker teams, located at the bottom of the figure), we can see prolonged periods with unchanging values of the strength value, which correspond to the absent observations. Similar figures for the static and random walk models are given in Figures~\ref{fig:figHockeyStrengthStatic} and \ref{fig:figHockeyStrengthRandomWalk}.

If the values of the explanatory variables at time $T+1$ are known, they can be plugged into \eqref{eq:gasModel} to obtain the values of the worth parameters at $T+1$. These can in turn serve to make one-step-ahead predictions or estimate the probabilities of specific rankings or ranking-based events. Applications in betting are straightforward. For instance, one can easily obtain the probability that a particular team will win a medal or that the podium will be occupied by a given list of teams.

An example is presented in Table~\ref{tab:HockeyPredictions}. Assuming that none of the Big Six countries host the upcoming championships, we calculated the future value of the team strength, $f_{i, T+1}$, and the associated rank prediction for the Big Six based on our estimates of the mean-reverting model. Even though the team strengths have the mean-reverting tendency, short-run predictions can differ from the unconditional mean substantially: even though the Big Six occupy the first six places according to both the predicted rankings for $T+1$ and the ultimate rankings in Table~\ref{tab:HockeyStrengths}, the rankings themselves are notably different.

Plugging the values of $f_{i, T+1}$ into \eqref{eq:lucePartProb} yields the estimated probability of a partial ordering of interest at $T+1$. For instance, the estimated probability of the partial ordering (Finland, Canada, Russia) -- the predicted podium outcome -- is 1.85 percent. This probability is low mainly because the predicted strengths happen to be quite similar across the first five teams. Analogously, we can obtain the probability of winning a gold medal, which is presented in the fourth column of Table~\ref{tab:HockeyPredictions}. Note that the winning probabilities are markedly different despite the similar team strengths. In practical applications, one might be interested in general ranking-based events, such as the probability that a team finishes on the podium; these probabilities can easily be obtained by combining suitable elementary events. An example is given in the last column of Table~\ref{tab:HockeyPredictions}.

\begin{table}
    \caption{One-step-ahead rank prediction and medal probabilities for the Big Six in the mean-reverting model. No home-ice advantage assumed.}
    \label{tab:HockeyPredictions}
    
\begin{tabular}[t]{>{\raggedright\arraybackslash}p{2.5cm}rrrr}
\toprule
Country & Strength & Predicted rank & P[gold medal] & P[podium position]\\
\midrule
Finland & $3.974$ & 1 & $0.235$ & $0.630$\\
Canada & $3.970$ & 2 & $0.234$ & $0.629$\\
Russia & $3.431$ & 3 & $0.137$ & $0.431$\\
Czechia & $3.415$ & 4 & $0.134$ & $0.426$\\
Sweden & $3.400$ & 5 & $0.133$ & $0.421$\\
United States & $2.086$ & 6 & $0.036$ & $0.128$\\
\bottomrule
\end{tabular}
\end{table}

\section{Discussion of other applications}
\label{sec:disc}

\subsection{Underlying index}

There is often an underlying index or score behind a ranking. For example, the \emph{Times Higher Education World University Rankings} are based on the score weighted over 13 individual indicators grouped into five categories -- industry income, international diversity, teaching, research, and citations. International rankings based on various indices such as the \emph{Global Competitiveness Index}, \emph{Bloomberg Innovation Index}, \emph{Human Development Index}, \emph{Climate Change Performance Index}, and \emph{Good Country Index} are compiled in a similar fashion. Naturally, an analysis of these rankings and indices is a popular subject of scientific research; e.g., \cite{Saisana2005} assess the robustness of country rankings, \cite{Paruolo2013} measure the importance of individual variables in composite indicators, and \cite{Varin2016} investigate the role of citation data in the ratings of scholarly journals.

The time aspect is inherent in these rankings, as they are typically compiled annually. \cite{Leckie2009} highlight the need for the prediction of ratings in the context of school choice based on league tables. They model the test scores of individual students nested in schools using the multilevel random-intercepts model. As they note, the main goal is to obtain relative ratings of schools rather than changes in the mean and variance over time. The conclusion is that there is a substantial uncertainty in test scores and their ability to forecast school performance is therefore very limited. Nevertheless, it may prove to be interesting to model the rankings of schools using our proposed model.

In general, rankings can be modelled directly -- by a model for permutations -- or indirectly -- by a model for the underlying index. If it is reasonable to assume that the indices of individual items are independent, the best option might be to model just the underlying index using a univariate model. In many real-life applications, the independence of the items' index values is questionable, as the items may interact in various ways or share a common pool of resources. When there is a potential relationship, our dynamic model for rankings might be more suitable, as it naturally captures dependence between the items. Furthermore, in the end, the reader is often only interested in the eventual rankings anyway, as they are more illustrative and attractive than the underlying indices.

\subsection{Repeated surveys}

A common way of obtaining ranking data is through a survey in which respondents are asked to rank items. Many surveys are repeated on several occasions, forming time-series data. For the statistical methodology dealing with repeated surveys, see \cite{Scott1974} and \cite{Steel2009}. By asking ranking questions in repeated surveys, we arrive at a time series of rankings. For example, customers of a retail shop may be periodically asked to rank products according to their preferences. In this case, the proposed dynamic ranking model could be a useful tool.

\subsection{Non-parametric efficiency analysis}

Another interesting application is modelling the rankings of decision making units (DMUs) obtained by non-parametric efficiency analysis, such as the \emph{data envelopment analysis (DEA)} pioneered by \cite{Charnes1978} and \cite{Banker1984}. Typically, DEA is applied in fields such as banking, health care, agriculture, transportation, and education to analyse the performance of banks, hospitals, farms, airlines, and schools, respectively \citep{Liu2013}. The goal of such analyses is to separate efficient and inefficient DMUs, assign efficiency scores to them, and determine their ranking. Many empirical papers also study the determinants of efficiency. The analysis is usually carried out by first obtaining the efficiency scores and then analysing them using regression in the second phase. \cite{Simar2007} (i) point out that a vast majority of these analyses ignore the inherent dependence between efficiency scores in their second phase, and (ii) develop bootstrap procedures to fix invalid inference. 

\cite{Simar2007} focus on the cross-sectional case in which dependence only occurs between the DMUs, not over time, which also greatly facilitates bootstrapping. The extension to panel data is not straightforward to say the least. Nevertheless, in many empirical studies, DMUs are observed annually, with the intention of both assessing the way efficiency evolved over time and providing a list of units that proved to be capable of sustaining efficiency over a long period. For this type of analysis, it may be beneficial to model the dynamics of DEA rankings using our model. If the long-term efficiency is of interest, it can be measured via the unconditional ranking. A major limitation of this approach is, however, the use of the Plackett-Luce distribution, as DEA rankings do not obey Luce's choice axiom. Other, more complex distributions on rankings could prove more appropriate here. For example, a richer dependence structure can be provided by Thurstone order statistics models based on the multivariate normal distribution (see \citealp{Thurstone1927} and \citealp{Yu2000}) or multivariate extreme value distributions (see \citealp{McFadden1978} and \citealp{Joe2001}). Note that the latter class contains the Plackett-Luce model as a special case. 

Modelling rankings instead of efficiency scores may also enhance the robustness with regard to method selection. For example, a novel DEA approach utilising the Chebyshev distance proposed by \cite{Hladik2019} offers alternative efficiency scores to the classical DEA models of \cite{Charnes1978} and \cite{Banker1984}, but it has been shown to produce the exact same ranking. Modelling rankings instead of efficiency scores thus eliminates differences between the two methods.

\section{Conclusion}
\label{sec:con}

Our new modelling approach brings two main features that have not been utilised in the analysis of time-varying rankings so far: (i) it allows a general autoregressive scheme for the process that governs the items' worth parameters, and (ii) new observations can update the worth parameters through a score-driven mechanism. Both of these features proved useful in our case study dealing with ice hockey team rankings. We believe that empiricists in diverse application areas can benefit from these features as well. These empiricists will hopefully also appreciate other practical merits of the model, such as the ability to include time-varying covariates or the straightforward maximum likelihood estimation.

This paper has presented the first results of ongoing research. Future efforts should mainly cover the following areas. First, we hope to see more complex results regarding both finite-sample performance and limit behavior. We doubt that comprehensive analytical treatment of the maximum likelihood asymptotics is tractable, but we aim to extend the current simulation results substantially. Second, for applications with a very large number of items, empiricists would surely benefit from a specialised algorithm for likelihood maximisation that exploits the specific structure of the likelihood function. As we mentioned above, \cite{Creal2013} and \cite{Caron2012} might provide useful inspiration in this respect. Finally, for applications to rankings where ties are possible, the model can be extended using the approach of \cite{Firth2019} and \cite{Turner2020}.

\section*{Acknowledgements}
\label{sec:acknow}

We would like to thank Michal Černý for his comments. Computational resources were supplied by the project `e-Infrastruktura CZ' (e-INFRA LM2018140) provided within the program Projects of Large Research, Development and Innovations Infrastructures.

\section*{Funding}
\label{sec:fund}

This research was supported by the Internal Grant Agency of Prague University of Economics and Business under project F4/27/2020 and the Czech Science Foundation under project 19-08985S.

\appendix

\section{Placket-Luce probabilities, log-likelihood, and score with three items}
\label{sec:threeItems}

We present an example of equations \eqref{eq:luceCompProb}, \eqref{eq:luceCompLik}, and \eqref{eq:luceCompScore} for the case of three items. 
The probability mass function is given by
\begin{equation}
\label{eq:luceThreeProb}
\PP \left[ Y = y \middle| f \right] = \frac{\myexp{f_{\nth{1}}}}{\myexp{f_{\nth{1}}} + \myexp{f_{\nth{2}}} + \myexp{f_{\nth{3}}}}\cdot \frac{\myexp{f_{\nth{2}}}}{\myexp{f_{\nth{2}}} + \myexp{f_{\nth{3}}}}.
\end{equation}
The log-likelihood function is given by
\begin{equation}
\label{eq:luceThreeLik}
\ell \left( f \middle| y \right) = f_{\nth{1}} + f_{\nth{2}} - \ln \left( \myexp{f_{\nth{1}}} + \myexp{f_{\nth{2}}} + \myexp{f_{\nth{3}}} \right) - \ln \left( \myexp{f_{\nth{2}}} + \myexp{f_{\nth{3}}} \right).
\end{equation}
The score function is given by
\begin{equation}
\label{eq:luceThreeScore}
\begin{aligned}
\nabla_{\nth{1}} \left( f \middle| y \right) &= 1 - \frac{\myexp{f_{\nth{1}}}}{\myexp{f_{\nth{1}}} + \myexp{f_{\nth{2}}} + \myexp{f_{\nth{3}}}}, \\
\nabla_{\nth{2}} \left( f \middle| y \right) &= 1 - \frac{\myexp{f_{\nth{2}}}}{\myexp{f_{\nth{1}}} + \myexp{f_{\nth{2}}} + \myexp{f_{\nth{3}}}} - \frac{\myexp{f_{\nth{2}}}}{\myexp{f_{\nth{2}}} + \myexp{f_{\nth{3}}}}, \\
\nabla_{\nth{3}} \left( f \middle| y \right) &= - \frac{\myexp{f_{\nth{3}}}}{\myexp{f_{\nth{1}}} + \myexp{f_{\nth{2}}} + \myexp{f_{\nth{3}}}} - \frac{\myexp{f_{\nth{3}}}}{\myexp{f_{\nth{2}}} + \myexp{f_{\nth{3}}}}. \\
\end{aligned}
\end{equation}
It is obvious that $\nabla_{\nth{1}}( f | y ) + \nabla_{\nth{2}}( f | y ) + \nabla_{\nth{3}}( f | y ) = 0$, as fractions in \eqref{eq:luceThreeScore} with the same denominator sum to one. Moreover, it is easily seen that $\nabla_{\nth{1}} \in (0, 1)$, $\nabla_{\nth{2}} \in (-1, 1)$, and $\nabla_{\nth{3}} \in (-2, 0)$, as each fraction has a value between 0 and 1. 

\section{Deriving properties of the Plackett-Luce model via the softmax and logsumexp functions}
\label{sec:softmax}

The analysis of the Plackett-Luce model is facilitated by the use of the softmax and logsumexp functions, which are denoted here as $\sigma(\cdot)$ and $\lse(\cdot)$, respectively. In this appendix, we first employ these functions to easily derive the shape of the likelihood and score, and then use them to study the properties of the score. 

Recall that for an $n$-vector $z$, $\sigma(z)_i = \myexp{(z_i)}/\sum_{j=1}^n \myexp{z_j}$ and $\lse(z) = \log \sum_{j=1}^n \myexp{z_j}$. It is easily verified that
\begin{align}
		\log\sigma(z)_i &= z_i - \lse(z), \label{eq:logSoftmax}\\
		\frac{\partial\lse(z)}{\partial z_i} &= \sigma(z)_i\,. \label{eq:diffLse}
\end{align}
To simplify formulas, we introduce the shorthand notation $f_{\geq r}$ for a vector containing worth parameters of items ranked $\rth{r}$ or worse, i.e., $f_{\geq r} = (f_i)_{\{i\in\mathcal{Y}: y^{-1}(i) \geq r\}}$. With this notation, we can rewrite \eqref{eq:luceCompProb} as
\begin{equation}
	\label{eq:luceProbSoftmax}
		\PP \left[ Y = y \middle| f \right] = \prod_{r=1}^{N} \sigma(f_{\geq r})_r.
\end{equation}
Combining \eqref{eq:logSoftmax} and \eqref{eq:luceProbSoftmax} immediately yields
\begin{equation}
	\label{eq:luceLoglikSoftmax}
	\ell \left( f \middle| y \right) = \sum_{i=1}^{N} f_i - \sum_{r=1}^{N} \lse({f_{\geq r}}),
\end{equation}
and using \eqref{eq:diffLse}, we obtain the score for player $i$ in the form
\begin{equation} 
	\label{eq:luceScoreSoftmax}	
	\nabla_i \left( f \middle| y \right) = 1 - \sum_{r = 1}^{y(i)} \sigma(f_{\geq r})_i\,.
\end{equation}

Since (i) values of the softmax function lie in the $(0,1)$ interval and (ii) $\sigma(f_{\geq N}) = 1$, we can easily establish the following bounds for the score. The score lies in $(1 - r, 1)$ for an item with rank $r = 1,\ldots,N-1$, and in $(1-N, 0)$ for the item ranked last (\rth{N}). The bounds are tight, as the following example with a dominant item demonstrates. Consider the identity ranking $y(i) = i$ and worth parameters 
$$
f_i = 
\begin{cases}
	c               & \text{if }       i = d, \\
	\frac{-c}{N-1}  & \text{otherwise},
\end{cases}
$$
where $c>0$ and $d \in \mathcal{Y}$ is a dominant item. (It is easily verified that $\sum_{i=1}^N f_i = 0$.) For $r\leq\min(i,d)$ we obtain
\begin{equation} 
	\label{eq:softmaxLimit}
	\lim_{c \to \infty}\sigma\big(f_{\geq r}\big)_i  = 
	\begin{cases}
		1               & \text{if }       i = d, \\
		0  & \text{otherwise}.
	\end{cases}
\end{equation}
Combining \eqref{eq:luceScoreSoftmax} and \eqref{eq:softmaxLimit} yields $\lim_{c \to \infty} \nabla_d \left( f \middle| y \right) = 1 - d,$ which demonstrates that the lower bound for the score is tight. Setting $d = N$ (the dominant item unexpectedly ranks last) yields $\lim_{c \to \infty} \nabla_i \left( f \middle| y \right) = 1$ for $i = 1,\ldots,N-1$, which demonstrates the tightness of the upper bound of the score for all but the last item. (The tightness of the upper bound for the last item can be shown using a similar example with an inferior item that ranks last, as expected.)


\end{document}